\documentclass[10pt]{article}

\usepackage{graphicx}
\usepackage[latin1]{inputenc}
\usepackage{enumerate}
\usepackage{amsthm}
\usepackage{amstext}
\usepackage{amsxtra}
\usepackage{amsmath} \usepackage{amsfonts} \usepackage{amssymb}

\begin{document}
\sloppy
\renewcommand{\theequation}{\arabic{section}.\arabic{equation}}
\thinmuskip = 0.5\thinmuskip    
\medmuskip = 0.5\medmuskip
\thickmuskip = 0.5\thickmuskip
\arraycolsep = 0.3\arraycolsep

\newtheorem{theorem}{Theorem}[section]
\newtheorem{corollary}[theorem]{Corollary}
\newtheorem{lemma}[theorem]{Lemma}
\newtheorem{proposition}[theorem]{Proposition}
\renewcommand{\thetheorem}{\arabic{section}.\arabic{theorem}}

\newcommand{\prf}{\noindent{\bf Proof.}\ }
\def\prfe{\hspace*{\fill} $\Box$

\smallskip \noindent}

\def\be{\begin{equation}}
\def\ee{\end{equation}}
\def\bea{\begin{eqnarray}}
\def\eea{\end{eqnarray}}
\def\beas{\begin{eqnarray*}}
\def\eeas{\end{eqnarray*}}

\newcommand{\R}{\mathbb R} 
\newcommand{\N}{\mathbb N}

\def\supp{\mathrm{supp}\,} 
\def\vol{\mathrm{vol}\,} 
\def\sign{\mathrm{sign}\,}

\title{Existence of axially symmetric solutions to the Vlasov-Poisson system depending on Jacobi's integral}
\author{ Achim Schulze\\
         Mathematisches Institut der
         Universit\"at Bayreuth\\
         D 95440 Bayreuth, Germany}
\maketitle

\begin{abstract}
We prove the existence of axially symmetric solutions to the Vlasov--Poisson system in a rotating setting for sufficiently small angular velocity. The constructed steady states depend on Jacobi's integral and the proof relies on an implicit function theorem for operators. 
\end{abstract}

\section{Introduction}
\setcounter{equation}{0}
In stellar dynamics, the evolution of a large ensemble of particles (e.g. stars) which interact only by their self-consistent, self-generated gravitational field, is described by the Vlasov-Poisson system
\be
\partial_{t}f+v\cdot \nabla_{x}f-\nabla_{x}U\cdot \nabla_{v}f =0,
\label{vlasov10}
\ee
\be
\Delta U = 4\pi \rho,  \label{poissonn}
\ee
\be
\rho(t,x) = \int f(t,x,v)dv \label{rhodef2}.  
\ee
Here $f= f(t,x,v) \geq 0$ is the phase-space density, where $t\in\R$ denotes time, and $x,v\in \R^3$ denote position and velocity. 
$U=U(t,x)$ is the gravitational potential of the ensemble, and $\rho=\rho(t,x)$
is its spatial density.
We are looking for stationary solutions of (\ref{vlasov10})-(\ref{rhodef2}). 
The ansatz
\begin{equation} \label{stdansatz}
f_0(x,v) = \Phi (E) = \Phi (\frac{1}{2} v^2 + U(x))
\end{equation} 
is well known and autmoatically satisfies the Vlasov equation (\ref{vlasov10}), because the particle energy
$$ E(x,v):= \frac{1}{2} v^2 + U(x) $$
is a conserved quantity along characteristics. But we still have to construct the self-consistent potential. This is done by plugging (\ref{stdansatz}) into the Poisson equation, more precisely, we have to solve
\begin{equation} \label{poissonsolveplug}
\Delta U = 4\pi h_{\Phi} (U) = 4\pi \int \Phi (\frac{1}{2} v^2 + U(x)) dv
\end{equation}
The ansatz (\ref{stdansatz}) only leads to spherically symmetric stationary solutions of (\ref{vlasov10})-(\ref{rhodef2}), where $f$ is called spherically symmetric, iff $f(Ax,Av) = f(x,v) \quad \forall \, \, A \in O(3)$. Indeed, this is a special case of a more general result of Gidas, Ni and Nirenberg, cf. \cite{GNN}. If one is interested in stationary solutions with less symmetry, more invariants can be added to (\ref{stdansatz}), so that the right-hand side of (\ref{poissonsolveplug}) explicitly depends on $x$.  \\ One possibility is to consider a rotating system. If the ensemble is rotating around a given axis, say the $x_3$-axis, we can change to the rotating frame and change coordinates as follows:
$$\zeta := R_t x, \qquad \eta :=  R_t v - \Omega \times (R_t  x), $$
where 
$$ R_t:= \left( \begin{array}{ccc} \cos(\omega t) & \sin(\omega t) & 0 \\  -\sin (\omega t) & \cos(\omega t) & 0 \\ 0 & 0 & 1 \end{array} \right), \quad \Omega := \left( \begin{array}{c} 0 \\ 0 \\ \omega \end{array} \right)$$
and the (rotational) velocity $\omega >0$ is given.
The Vlasov-Poisson system then takes the form
\be
\partial_{t}f+\eta \cdot \nabla_{\zeta}f-\left( \nabla_{\zeta}U + \Omega \times (\Omega \times \zeta) + 2(\Omega \times \eta) \right) \cdot \nabla_{\eta}f =0,
\label{vlasov2}
\ee
\be
\Delta_{\zeta} U(t,\zeta) = 4\pi \rho (t,\zeta), \label{poisson20}
\ee
\be
\rho(t,\zeta) = \int f(t,\zeta,\eta)d\eta \label{rhodef20}
\ee
and the characteristic system of the Vlasov equation (\ref{vlasov2}) reads 
$$ \left\{ \begin{array} {lll} \dot{\zeta} & = & \eta \\ \dot{\eta} & = & -\partial _{\zeta} U(t,\zeta) - 2 \Omega \times \eta - \Omega \times (\Omega \times \zeta) \end{array} \right. , $$ 
which has the following expression as a conserved quantity, if $U$ is time-independent:
$$ E_J := \frac{1}{2} \eta ^2 + U (\zeta) - \frac{1}{2} |\Omega \times \zeta |^2,$$
where $E_J$ is also called Jacobi's integral.
A natural ansatz for the construction of stationary solutions of (\ref{vlasov2})-(\ref{rhodef20}) is now
\begin{equation}
f(\zeta ,\eta ) = \varphi (E_J) = \varphi( \frac{1}{2} |\eta|^2 + U(\zeta) - \frac{1}{2} \omega ^2 r^2 )
\end{equation} for a suitable function $\varphi : \mathbb{R} \rightarrow \mathbb{R}^+$, where $r:=r(x)= \sqrt{\zeta_1^2+\zeta_2^2}$.
In the original coordinates $x,v$ one easily verfies that this ansatz leads to
$$ g(x,v) := f(\zeta, \eta) = \varphi( \frac{1}{2} v^2 + U(R_t x) -  \omega P),$$
where we define $P$ as the third component of the angular momentum, that is $P:= x_1 v_2 - x_2 v_1$, which is a conserved quantity of the characteristic system of the Vlasov equation (\ref{vlasov10}), if $U$ is axially symmetric with respect to the $x_3$-axis.
Obviously, the function $f= f(\zeta,\eta)$ then automatically satisfies (\ref{vlasov2}) and one has to solve the Poisson equation, where we relabel $\zeta$ and $\eta$ to $x$ and $v$,\\ 
\begin{equation}
\Delta U = \int \varphi ( \frac{1}{2} v^2 + U(x) - \frac{1}{2}\omega ^2 r^2) \, dv =: \tilde{h} (\omega, r(x), U(x)).
\label{poissona}
\end{equation}
So if we construct an axially symmetric $U$ solving (\ref{poissona}), the corresponding functions $(g,U)$, with $g$  defined as above also will be a stationary solution of (\ref{vlasov10})-(\ref{rhodef2}). Clearly, our ansatz for $f$ satisfies (\ref{vlasov2}) without any
symmetry assumptions on $U$ and this gives hope for the construction of stationary solutions with less symmetry, for example triaxial systems. \\
Equation (\ref{poissona}) has been studied, among others, by Vandervoort, cf. \cite{VAN}. He observed numerically, 
that if $\varphi$ is of the form 
\be \label{betasol}
\varphi(E_J) = (E_0 -E_J)_+^{\beta -3/2},
\ee 
then for $0.5 < \beta \leq 0.808$ there are triaxial solutions to (\ref{poissona}) for sufficiently large $\omega$. For small $\omega$ or $\beta>0.808$, all numerically constructed solutions are axially symmetric. 
Consequently, (\ref{poissona}) seems to be of particular interest for the construction of ellipsoidal systems, but to our knowledge no self-consistent
ellipsoidal systems to (\ref{vlasov10})-(\ref{rhodef2}) or (\ref{vlasov2})-(\ref{rhodef20}) have been constructed analytically yet. \\
We will prove that there exist axially symmetric solutions to (\ref{poissona}) for small $\omega$ under suitable assumptions on $\varphi$, where we treat the case $\beta > 5/2$ in (\ref{betasol}).
For this purpose, we require, that for $\omega = 0$, we have a nontrivial, spherically symmetric solution $(f_0,U_0)$ of (\ref{poissona}). Note, that in this case the righthand-side of (\ref{poissona}) only depends on $U_0$. For $\omega \neq 0$, we want to apply an implicit function theorem to get solutions, which arise by deforming $U_0$, where certain symmetries are conserved. The central idea, which makes this approach work is to look for a solution $U^{\omega}$ as a deformation of $U_0$, i.e., $U^{\omega} = U_0(g(x))$ for some diffeomorphism $g$ on $\mathbb{R}^3$, and to formulate the problem in terms of finding zeros of a suitable operator $T$ over the space of such deformations instead of the space of the potentials. Whereas the original problem (\ref{poissona}) had to be solved in $\mathbb{R}^3$, we will only need to know the deformation on a compact neighbourhood of the support of the original solution $(f_0,\rho _0, U_0)$, and this provides useful compactness properties. Furthermore, finite radius and finite mass of the constructed solutions then are just consequences of the corresponding properties of $(f_0,\rho _0, U_0)$.\\
Although the allowed perturbations for the potential $U_0$ only have mirror symmetry which would match a triaxial system, we have up to now no method to exclude axial symmetry with respect to the $x_3$-axis for the perturbations constructed by the implicit function theorem. \\
The approach described above has been used by Lichtenstein for proving the existence of slowly rotating Newtonian stars, as described by selfgravitating fluid balls, cf. \cite{LICHT,LICHTA}. A translation of Lichtenstein's approach into modern mathematical language is due to Heilig, cf. \cite{HEILIG}. \\ The investigations made there were applied to the Vlasov-Poisson system in \cite{REIN}, where stationary solutions to (\ref{vlasov10})-(\ref{rhodef2}) of the form $f(x,v) = \varphi(E)\psi(\omega P)$ were constructed. There, the potential $U$ a-priori was axially symmetric, so that the expression $P = x_1v_2 - x_2v_1$ is a conserved quantity with respect to the characteristic system. The procedure described there is the basis of our approach. \\
This paper is organized as follows: In the next section we rewrite the problem in terms of finding zeros of the operator $T$, we then state the main result and prove it using an implicit function theorem. For this, we need certain properties of $T$ which can be proved as in \cite{REIN}, except some minor technical modifications and one lemma, where the symmetry of the allowed perturbations enters in. In Section \ref{dT2}, we generalize this important lemma dealing with properties of the operator $\partial _{\zeta} T(0,0)$ to mirror symmetry.\\ 
\section{The main result}
The mappings, which leave our solutions invariant, are in the set 
\begin{align*}
S:= \{ &\tau_{110} : (x_1,x_2,x_3) \mapsto (x_1,x_2,-x_3), \enspace \tau_{101} : (x_1,x_2,x_3) \mapsto (x_1,-x_2,x_3), \\
 \enspace &\tau_{011} : (x_1,x_2,x_3) \mapsto (-x_1,x_2,x_3)\}.
\end{align*}
Now let $B_R := \{ x \in \mathbb{R}^3 \, | \, |x| \leq R \}$ and define
\begin{equation}
C_S (B_R) := \{ f \in C(B_R) | \enspace f(Ax) = f(x), \enspace A \in S, \enspace x \in B_R \}.
\end{equation}
Then we have
$$ \nabla f(0) = 0, \quad \text{if} \quad f \in C^1 (B_R) \cap C_S (B_R) .$$
For $\varphi  : \mathbb{R} \rightarrow [0, \infty[ $ we require
\begin{itemize}
\item[$(\varphi 1)$] $\varphi \in C^1 (\mathbb{R})$ and there is $E_0 \in \mathbb{R}$ with $\varphi(E_J) = 0$ for $E_J\geq E_0$ and $\varphi(E_J) > 0$ for $E_J<E_0$. 
\item[$(\varphi 2)$] $\varphi$ is strictly decreasing in $]-\infty, E_0[$.
\item[$(\varphi 3)$] The ansatz $f_0 (x,v) = \varphi(E_J)$ with $\omega =0$ produces a nontrivial, spherically symmetric solution $(f_0,\rho _0, U_0)$ of (\ref{vlasov10})-(\ref{rhodef2}) with
$\rho _0 \in C^1_0 (\mathbb{R}^3)$, supp $\rho _0 = B_1$ and $U_0 \in C^2 (\mathbb{R}^3)$ with $\lim_{|x| \rightarrow \infty} U_0(x) = 0$. 
\end{itemize}
Examples for a functions satisfying $(\varphi 1)$--$(\varphi 3)$ are the so-called polytropes
$$ \varphi (E_J) := (E_0 - E_J)_+^k $$
for $k>1$ and suitable $E_0 < 0$.
Now we can state the main theorem.
\begin{theorem}  \label{mainth}
Let $r:= \sqrt{x_1^2+x_2^2}$. There exists $\omega _0 >0$, such that for all $\omega \in ]-\omega _0, \omega _0[$ there exits a nontrivial solution $(f^{\omega}, \rho ^{\omega}, U^{\omega})$ of (\ref{vlasov2})-(\ref{rhodef20}) with
\begin{itemize}
\item[(i)] $f^{\omega} (x,v) = \begin{cases} \varphi (\frac{1}{2} v^2 + U^{\omega} (x) - \frac{1}{2}\omega^2r^2 ) &\text{for} \quad |x| < 4 \\ 0 &\text{else} \end{cases}$
\item[(ii)] $(f^0, \rho ^0, U^0) = (f_0, \rho _0, U_0)$ and for $|\omega | < \omega _0$, $(f^{\omega}, \rho ^{\omega}, U^{\omega})$ has the following symmetry properties: For all $A \in S$ we have
$$ f^{\omega} (Ax,Av) = f^{\omega} (x,v), \quad \rho^{\omega} (Ax) = \rho(x), \quad U^{\omega} (Ax) = U^{\omega} (x) $$
and $(f^{\omega}, \rho ^{\omega}, U^{\omega})$ is not spherically symmetric for $\omega \neq 0$.
\item[(iii)] $\rho ^{\omega} \in C^1_c (\mathbb{R}^3)$ and $U^{\omega} \in C^2_b (\mathbb{R}^3)$, where $\rho ^{\omega} (x) = \int f^{\omega} (x,v) \, dv$.
\item[(iv)] The mappings $]-\omega _0, \omega _0[ \ni \omega \mapsto \rho ^{\omega}$ and $]-\omega _0, \omega _0[ \ni \omega \mapsto U^{\omega}$ are continuous with respect to the norms $\| \cdot \|_{1,\infty}$ or
$\| \cdot \|_{2,\infty}$, respectively.
\end{itemize}
\end{theorem}
\noindent
\textbf{Remark.} If we add rotations about the $x_3$-axis to the set $S$, the proof of Theorem \ref{mainth} still holds -- we can essentially follow the proof given here, and this shows that the constructed solutions in Theorem \ref{mainth} have to be axially symmetric a-posteriori. This follows by the uniqueness of the mapping given by the implicit function theorem, cf. Theorem \ref{implicit}. \\ \\
\noindent
For the proof of Theorem \ref{mainth}, we need some lemmata.
\begin{lemma} \label{lemmaueig}
The spherically symmetric solution $(f_0,\rho _0, U_0)$ has the following properties.
\begin{itemize}
\item[(a)] The potential $U_0$ is given by
$$ U_0(x) = - \int \frac{\rho _0 (y)}{|x-y|} dy = - \frac{4\pi}{|x|} \int_0^{|x|} s^2 \rho _0 (s)ds - 4\pi \int_{|x|}^{\infty} s \rho _0 (s) ds, \enspace x\in \mathbb{R}^3. $$
\item[(b)] $\rho _0$ is decreasing with $\rho _0 (0)>0, \, U_0 '' (0) >0$ and for every $R>0$ there exists $C>0$, such that $U_0'(r) \geq Cr, \quad r\in [0,R],$ and $U_0 (1) = E_0$.
\item[(c)] $\rho _0 '$ is H\"older continuous and $U_0 ' \in C^2 (\dot{\mathbb{R}}^3)$, where $\dot{\mathbb{R}}^3 := \mathbb{R}^3 \backslash \{0\} $. 
\end{itemize}
\end{lemma}
\begin{proof}
The formula
$$U_0'(r)= \frac{4\pi \int_0^r s^2\rho_0(s) \, ds}{r^2}$$
easily follows from the Poisson equation with spherical symmetry and since we require $\lim_{|x| \rightarrow \infty} U_0(x) = 0$, the representation for $U_0$ holds by uniqueness. As to (b), for $\omega =0$ we have $f_0(x,v) = f_0(E) = f_0(\frac{1}{2} v^2 + U_0(x))$ and this implies
\be \label{rhoformel}
\rho _0 (x) = \int_{\mathbb{R}^3} f_0(x,v) \, dv = h_0(U_0(x)) := 4\pi \sqrt{2} \int_{U_0(x)}^{E_0} \varphi (E) \sqrt{E-U_0(x)}  \, dE,
\ee
where the function $h$ is continuously differentiable and with $(\varphi1),\, (\varphi2)$ we have $h'(s)<0$ for $s < E_0$. Consequently, $\rho _0$ is decreasing
because $U_0$ is increasing and since the steady state $(f_0,U_0)$ is assumed to be nontrivial, we must have $\rho_0(0) >0$. Thus actually $U_0'(r)>0$, $r>0$, and since $U_0''(0) = (4\pi/3)\rho_0(0) >0$ this implies the estimate on $U_0'$ from below. The assertion that $U_0(1) =E_0$ follows from (\ref{rhoformel}) and the assumption $\supp \rho_0 =B_1$. The regularity of $U_0'$ follows from the formula for $U_0'$ above and the fact that $\rho_0 \in C^1_c$, which we deduce again from (\ref{rhoformel}). Finally, the H\"older continuity of $\rho_0'$ will be part of the next Lemma.
\end{proof}
\begin{lemma} \label{lemma21}
Let  $E_1:= U_0(2)-E_0$ and define $f$ by $$f(x,v) =  \begin{cases} \varphi (\frac{1}{2} v^2 + U (x) - \frac{1}{2} \omega^2 r^2) &\text{for} \quad U(x)< E_0 +E_1  \\ 0 &\text{else} \end{cases},$$\\ where $\varphi$ satisfies
$(\varphi 1)$, $(\varphi 2)$ and $U \in C^2_b (\mathbb{R}^3)$ with $U (x) > E_0 + E_1$ for $|x| > 4$. Then the following holds:
\begin{align} \label{tildeh}
\rho_f (x) &:= \int_{\mathbb{R}^3} f(x,v) \, dv \notag \\
&= \tilde{h} (\omega , r(x), U(x)) \notag \\
&= \begin{cases} h(U(x) - \frac{1}{2} \omega ^2 r^2) &\text{for} \quad U(x)< E_0 +E_1 \\ 0 &\text{else} \end{cases} 
\end{align}
with 
$$ h(s) = 4\pi \sqrt{2} \int_s^{E_0} \sqrt{E-s}\, \varphi (E) dE .$$
Furthermore, $\tilde{h} \in C^1 (\mathbb{R} \times [0,\infty[ \times \mathbb{R} )$
and for every bounded set $B \subset \mathbb{R} \times [0,\infty[ \times \mathbb{R}$ there are constants $C>0$ and $\mu \in ]0,1[$ such that for
$ (\omega , r ,u), (\omega ', r , u') \in B$ we have
\begin{align*}
&|\partial _r \tilde{h} (\omega , r, u) | \leq Cr ,\\
&|\tilde{h} (\omega , r, u) - \tilde{h} (\omega ' , r, u') | \leq C( |\omega - \omega '|r + |u-u'|) ,\\
&|\partial _u \tilde{h} (\omega , r, u) - \partial _u \tilde{h} (\omega ' , r, u') | \leq C ( |\omega - \omega '| + |u-u'|^{\mu}).
\end{align*}
In addition, for $\omega = 0$, the function $\tilde{h} (0,\cdot,\cdot)$ does not depend on $r(x)$ and we can write $h_0 := \tilde{h}(0,0,u)$.
\end{lemma}
\begin{proof}
Introducing polar coordinates, we have for $U(x) < E_0+E_1$
\begin{align*}
\rho (x) &= \int \varphi \left(\frac{1}{2} v^2 + U(x) - \frac{1}{2} \omega^2 r^2\right)dv \\ 
&= 4\pi \int_0^{\infty} t^2 \varphi \left(\frac{1}{2} t^2 + U(x) -\frac{1}{2} \omega ^2 r^2 \right) dt \\
&= 4\pi \sqrt{2} \int_{U(x)-\frac{1}{2}\omega^2r^2}^{E_0} \left( E-U(x)+\frac{1}{2}\omega^2r^2 \right)^{1/2} \varphi(E) \, dE,
\end{align*}
and (\ref{tildeh}) follows. \\ We have $h \in C^1(\mathbb{R})$ with
\begin{equation*}
h'(s) = - 4\pi \sqrt{2} \int_s ^{E_0} \frac{1}{2\sqrt{E-s}} \varphi (E) dE
\end{equation*}
for $s<E_0$ and $h'(s) = 0$ for $s \geq E_0$ and the first two estimates follow. Next,
\begin{align*}
h''(s) &= - 4\pi \sqrt{2}\frac{d}{ds} \int_0 ^{E_0-s} \frac{1}{2\sqrt{E}} \varphi (E+s) dE \\
&= -4\pi \sqrt{2} \int_0 ^{E_0-s} \frac{1}{2\sqrt{E}} \varphi '(E+s) dE \\ &= -4\pi \sqrt{2} \int_s ^{E_0} \frac{1}{2\sqrt{E-s}} \varphi '(E) dE
\end{align*}
yields local Lipschitz continuity of $\partial _u \tilde{h}$ with respect to $\omega$ and $u$
and the proof is complete.
\end{proof}
We want to find solutions of the equation
\begin{equation} \label{poissonsol}
\Delta U = 4\pi \tilde{h} (\omega, r(x), U)
\end{equation}
and the main idea is to rewrite problem (\ref{poissonsol}) in terms of finding zeros of an operator $T$, which does not act directly on the space of potentials, but on deformations of the given spherically symmetric potential $U_0$. We define Banach spaces, which will serve as domain and range of $T$ 
\begin{align*}
X := \{ &f \in C_S (B_4) | f(0) = 0 , f \in C^1 (\dot{B}_4), \exists C>0: |\nabla f(x)| \leq C, x \in \dot{B}_4, \\ &\forall x \in \partial B_1 : \lim_{t \rightarrow 0, t>0} \nabla f(tx) =: \nabla f(0x) \enspace \text{exists, uniformly in} \enspace x \in \partial B_1\},
\end{align*}
where $\partial B_1 := \{x \in \mathbb{R}^3 \enspace | \enspace |x| =1\}$ and $\dot{B}_4 := B_4 \backslash \{0\}$. We equip $X$ with the norm
$$ \| f \|_X := \sup_{x \in \dot{B}_4} |\nabla f(x)|, \quad f \in X$$
and
\begin{align*}
Y := \{ &f \in C_S (B_4) | f(0) = 0 , f \in C^1 (B_4), \exists C>0: |\nabla f(x)| \leq C|x|, x \in B_4, \\ &\forall x \in \partial B_1 : \lim_{t \rightarrow 0, t>0} \frac{\nabla f(tx)}{t} =: \frac{\nabla f(0x)}{0} \enspace \text{exists, uniformly in} \enspace  x \in \partial B_1\}
\end{align*}
with norm
$$ \| f \|_Y := \sup_{x \in \dot{B}_4} \frac{ |\nabla f(x)|}{|x|}, \quad f \in Y.$$
To state more precisely, how to use functions in $X$ to deform the potential $U_0$, we need the next lemma.
\begin{lemma} \label{deformation}
For $\zeta \in X$ let
$$ g_{\zeta}: B_4 \rightarrow \mathbb{R}^3, \quad g_{\zeta} (x):= x + \zeta (x) \frac{x}{|x|}, \quad x\in \dot{B}_4, \quad  g_{\zeta} (0)=0 $$
Then there exists $r>0$, such that for all $\zeta \in \Omega$, where
$$ \Omega := \{ \zeta \in X | \|\zeta \|_X < r \} $$
we have:
\begin{itemize}
\item[(a)]  $g_{\zeta}: B_4 \rightarrow B_{4,\zeta}:=  g_{\zeta} (B_4)$ is a homeomorphism,  $g_{\zeta}: \dot{B}_4 \rightarrow \dot{B}_{4,\zeta}$ is a $C^1$-diffeomorphism, with
$$ |Dg_{\zeta} (x) - \text{id}| < \frac{1}{2}, \quad x \in \dot{B}_4 $$
and for every $x \in \partial B_1$ the mapping
$$ g_{\zeta} : \overline{0,4x} \ni y \mapsto g_{\zeta} (y) \in \overline{0,|g_{\zeta}(4x)|x} $$
is one-to-one, onto and preserves the natural ordering of points in $\overline{0,4x}$, where we defined $\overline{x_1,x_2} := \big\{ x_1 + \lambda (x_2-x_1) \, | \, \lambda \in [0,1] \big\}$ for $x_1,x_2 \in \mathbb{R}^3$.
\item[(b)] $\frac{1}{2}|x| \leq |g_{\zeta}(x)| \leq \frac{3}{2} |x|, \, x \in B_4$, and $g_{\zeta}(B_2) \subset \mathring{B}_3, \, B_3 \subset g_{\zeta} (B_4) \subset B_5$
\item[(c)] $g_{\zeta}(Ax) = A g_{\zeta} (x), \, x \in B_4$ and $g_{\zeta}^{-1} (Ax) = Ag_{\zeta}^{-1} (x), \, x \in B_{4,\zeta} ,\, A \in S$
\item[(d)] $|Dg_{\zeta}^{-1}(x) - \text{id}| < \frac{1}{2}, \, x \in \dot{B}_{4,\zeta}$ and there exists a constant $C>0$, such that for all $\zeta, \zeta ' \in \Omega$:
$$\frac{1}{|x|} |g_{\zeta} (x) - g_{\zeta '} (x)| + |Dg_{\zeta}(x) - Dg_{\zeta '} (x)| \leq C \| \zeta - \zeta '\|_X, \quad x \in \dot{B}_4, $$
and
$$ |g_{\zeta}^{-1} (x) - g_{\zeta '}^{-1} (x) | \leq C \| \zeta - \zeta '\|_X |x|, \quad x \in B_3 $$
\end{itemize}
\end{lemma}
\begin{proof}
In $\dot{B}_4$, we have for $i,j=1,2,3$:
\begin{equation} \label{dzeta}
\partial _{x_i} g_{\zeta,j} (x) = \delta _{ij} + \partial _{x_i} \zeta (x) \frac{x_j}{|x|} + \frac{\zeta (x)}{|x|} \left( \delta _{ij} - \frac{ x_i x_j}{|x|^2} \right)
\end{equation}
and therefore
$$  |Dg_{\zeta} (x) - \text{id}| < 3\|\zeta \|_X .$$
With the inverse function theorem the first two assertions in (a) follow.
For $x \in \partial B_1$,
$$ g_{\zeta} (tx) = tx + \zeta (tx) x = x \left( t + \zeta (tx) \right) $$
and
$$ \frac{d}{dt} \left( t + \zeta (tx) \right) = 1+ \nabla \zeta (tx) \cdot x > 0  \quad \text{for} \, \|\zeta \|_X \, \text{small} $$
and the proof of (a) is complete. \\
We have $| \zeta (x) | \leq \| \zeta \|_X |x|$ for $x\in B_4$ and this implies (b) for $r>0$ sufficiently small.  Assertion (c) is easily verified, too. If we choose $r$ even smaller we also have the first claim of (d), because
$$ D g_{\zeta}^{-1} (x) = (Dg_{\zeta})^{-1} (g_{\zeta}^{-1} (x)).  $$
The estimate for $g_{\zeta} - g_{\zeta '}$ follows from the definition of $g_{\zeta}$ and the estimate for $Dg_{\zeta} - Dg_{\zeta '}$ follows from (\ref{dzeta}).  \\
For $x \in \dot{B}_3$, we have with (b): $x \in g_{\zeta} (B_4) \cap  g_{\zeta '} (B_4)$. Consequently, there exists
$ y \in \dot{B}_4$ mit $x= g_{\zeta '} (y)$. Now we have
\begin{align*}
|g_{\zeta}^{-1} (x)- g_{\zeta '}^{-1} (x)| &= |g_{\zeta}^{-1} (g_{\zeta '} (y)) - y| \\ 
&= |g_{\zeta}^{-1} (g_{\zeta '} (y)) - g_{\zeta}^{-1} (g_{\zeta } (y))| \\
&\leq 2|g_{\zeta} (y) - g_{\zeta '} (y)| \leq 2 \|\zeta - \zeta '\|_X |y| \\
&\leq 4 \|\zeta - \zeta '\|_X |x|,
\end{align*}
where we used the mean value theorem, the estimate for $Dg_{\zeta}^{-1}$ and
$\overline{g_{\zeta}(y),g_{\zeta '}(y)} \subset g_{\zeta} (\dot{B}_4) $.
\end{proof}
We want to find solutions of (\ref{poissonsol}) with the following structure
$$ U(x) = U_{\zeta} (x) := U_0 (g_{\zeta}^{-1} (x)), \quad x \in B_{4,\zeta}, $$
with a suitable $\zeta \in \Omega$. Obviously, we need $U$ on the whole space $\mathbb{R}^3$, but this is only a technical problem. We use the fundamental solution of the Poisson equation to integrate (\ref{poissonsol}) and we then have to solve
\begin{equation} \label{opequ}
U_0 (x) + \int_{B_{4,\zeta}} \frac{\tilde{h} (\omega, r(y), U_0 (g_{\zeta}^{-1} (y)))}{|g_{\zeta} (x) -y|} dy = 0, \quad x \in B_4.
\end{equation}
This equation essentially contains the operator we are looking for, but we have to modify things a little and also we want to get rid of the dependence on $\zeta$ in the integration domain.
\begin{proof}[Proof of Theorem \ref{mainth}]
For $\zeta \in \Omega$ and $\omega \in \mathbb{R}$, we define
\begin{align} \label{opdefi}
T(\omega, \zeta) (x) &:= U_0 (x) + \int_{B_3} \frac{\tilde{h} (\omega, r(y), U_0 (g_{\zeta}^{-1} (y)))}{|g_{\zeta} (x) -y|} dy \notag \\ &- U_0 (0) - \int_{B_3} \frac{\tilde{h} (\omega, r(y), U_0 (g_{\zeta}^{-1} (y)))}{|y|} dy, \qquad x \in B_4.
\end{align}
Suppose we already know that this defines a continuous operator $$T: ]-\tilde{\omega},\tilde{\omega}[\times \Omega \rightarrow Y$$ for some $\tilde{\omega} >0$  and $T$ is continuously Frechet-differentiable with respect to $\zeta$, where
$$ \partial _{\zeta} T(0,0): X \rightarrow Y $$
is an isomorphism -- the first two assertion follow from \cite{REIN}, Section 2 and the last assertion will be verified here in Section \ref{dT2}. It is also there that the symmetry of the perturbations plays a crucial role. \\The definition of $Y$ requires $T(\omega, \zeta) (0) =0$ and therefore we substracted the constant in (\ref{opdefi}). With assumption ($\varphi 3$), we know $T(0,0)=0$, because $g_0 =$ id and $\text{supp} \, \rho _0 = \text{supp} \, h_0 \circ U_0 = B_1 \subset B_3.$ The implicit function theorem, cf. \cite{DEIMLING}, Theorem 15.1, also stated in the Appendix as Theorem \ref{implicit}, cf. Section \ref{appendix}, now guarantees the existence of  $\omega _1 \in ]0,\tilde{\omega}[$ and the existence of a continuous mapping
$$ ]-\omega _1, \omega _1[ \ni \omega \mapsto \zeta ^{\omega} \in \Omega $$
such that
$$ T(\omega,\zeta^{\omega}) = 0 , \qquad \omega \in ]-\omega _1, \omega _1[ $$
and $\zeta ^0 = 0$.
We also will require that $\omega ^2 r^2 < E_1$ in $B_4$, where $E_1$ is defined in Lemma \ref{lemma21} and therefore define
\be \label{defomega0}
\omega _0 := \min \bigg\{ \omega _1,\frac{ \sqrt{|E_1|}}{4}\bigg\} .
\ee
Now let $\zeta = \zeta^{\omega}$, where we choose a fixed $ \omega \in ]-\omega _0, \omega _0[$ and define
\begin{equation} \label{rhozeta}
\rho _{\zeta} (x) := \tilde{h} (\omega, r(x), U_0 (g_{\zeta} ^{-1} (x))), \qquad x \in B_3.
\end{equation}
Then we have $\rho _{\zeta} \in C_S (B_3) \cap C^1 (\dot{B}_3) $. By Lemma \ref{lemma21},
$\rho _{\zeta} >0 $ at most, if $U_0 (g_{\zeta} ^{-1} (x)) < E_0 + E_1$, which is equivalent to $|g_{\zeta}^{-1} (x)| < 2$ by Lemma \ref{lemmaueig}. Consequently,
$$ \text{supp} \rho_{\zeta} = g_{\zeta} (B_2) \subset \mathring{B}_3 .$$
We extend $\rho _{\zeta}$ by $0$ to all of $\mathbb{R}^3$ and we achieve
$$ \rho _{\zeta} \in C_c (\mathbb{R}^3), \quad \text{supp} \rho_{\zeta} \subset \mathring{B}_3.$$
We want equation (\ref{rhozeta}) to hold everywhere, but we have not defined $g_{\zeta}$ globally. \\
We can rewrite $T(\omega, \zeta) = 0$ as
$$ U_0(x) = - \int_{B_3} \frac{\rho _{\zeta} (y)}{|g_{\zeta}(x) -y|} dy + C, \quad x \in B_4,$$
or
$$ U_0 (g_{\zeta}^{-1} (x)) = - \int_{B_3} \frac{\rho _{\zeta} (y)}{|x -y|} dy + C, \quad x \in B_{4,\zeta},$$ 
where
$$ C:= U_0 (0) + \int_{B_3}  \frac{\rho _{\zeta} (y)}{|y|} dy .$$
Now define
$$ U_{\zeta} (x):= - \int_{\mathbb{R}^3}  \frac{\rho _{\zeta} (y)}{|x -y|} dy + C. $$
Then we have $U_{\zeta} \in C^1(\mathbb{R}^3)$ with
\begin{equation} \label{uzetainv}
U_{\zeta} (x) = U_0 (g_{\zeta}^{-1} (x)), \quad x \in B_3 \subset B_{4,\zeta}
\end{equation}
and thus $\rho _{\zeta} \in C^1_c (\mathbb{R}^3)$ and $U_{\zeta} \in C^2_b (\mathbb{R}^3)$ with $\Delta U_{\zeta} = 4\pi \rho _{\zeta} $ in $\mathbb{R}^3$. \\
Furthermore,
\begin{equation} \label{uzeta2}
\Delta U_{\zeta} = 4\pi \tilde{h} (\omega, r(x), U_{\zeta} (x)), \qquad x \in B_3 \subset B_{4,\zeta}.
\end{equation}
The last equation holds even in $\mathbb{R}^3$. We have to show
$$ \rho_{\zeta} (x) =  \tilde{h} (\omega, r(x), U_{\zeta} (x)), \quad x \in \mathbb{R}^3, $$
that is, $U_{\zeta} (x) > E_0 + E_1$ for $x \in \mathbb{R}^3 \backslash g_{\zeta} (B_2)$. We know
$$ \Delta U_{\zeta} (x) = 0, \qquad x\in  \mathbb{R}^3 \backslash g_{\zeta} (B_2),$$
$\lim_{|x| \rightarrow \infty } U_{\zeta} (x) = C$ and
\begin{align*}
& U_{\zeta} (x) = E_0 + E_1 , \qquad x \in \partial g_{\zeta} (B_2), \\
&U_{\zeta} (x) > E_0 + E_1 , \qquad x \in B_3 \backslash  g_{\zeta} (B_2).
\end{align*}
Here we used (\ref{uzetainv}) and the monotonicity of $U_0 (|x|)$ with $U_0(2)= E_0+E_1$.
If $C \leq E_0+E_1$, we have a contradiction to the maximum principle. Therefore,
$C> E_0 + E_1$ and again by the maximum principle: $U_{\zeta} > E_0+E_1$ on $\mathbb{R}^3 \backslash g_{\zeta} (B_2)$ and  consequently, (\ref{uzeta2}) holds in $\mathbb{R}^3$. \\ \\
Now define $\rho ^{\omega} := \rho _{\zeta}, \quad U^{\omega} := U_{\zeta}$ and
\begin{align}
f^{\omega} (x,v)&:= \begin{cases} \varphi (\frac{1}{2} v^2 + U^{\omega} (x) - \frac{1}{2} \omega^2 r^2),\quad &\text{for} \,\, U^{\omega} (x) < E_0 + E_1 \\ 0 &\text{else} \end{cases} \notag \\
&=  \begin{cases} \varphi (\frac{1}{2} v^2 + U^{\omega} (x) - \frac{1}{2} \omega ^2 r^2),\quad &\text{for} \, \, |x| < 4 \\ 0 &\text{else}. \label{fdefine} \end{cases} 
\end{align}
Now $f^{\omega}$ defined by (\ref{fdefine}) solves the Vlasov equation (\ref{vlasov2}) because it is constant along characteristics. More precisely, we have 
$U_\zeta (x) - \frac{1}{2} \omega^2 r^2 > E_0$ in a neighbourhood of $\partial B_4$, 
if we choose $\omega _0$ sufficiently small as in (\ref{defomega0}). 
If we then fix $(x,v)$ with $E_J(x,v) < E_0$ and consider a characteristic $(X,V)$ going through $(x,v)$ we conclude that if $x \in B_4$, we have $X \in B_4$ for all time. On the other hand, if $x \notin B_4$, we have $X \notin B_4$ for all time. \\
Altogether, assertions (i)-(iii) of the theorem follow, except the non-spherical symmetry in the case $\omega \neq 0$. Choose $x \in \mathbb{R}^3$ with $\rho^{\omega} (x) >0, \, x_1:= a \neq 0, x_2=x_3=0$. Then there exists some $\eta \in \mathbb{R}^3$, such that 
$$\frac{1}{2} \eta^2 + U^{\omega} (x) - \frac{1}{2} \omega ^2 a^2 < E_0.$$
Now if $(f^{\omega},U^{\omega})$ were spherically symmetric, there would exist a rotation $A$ around the $x_2$-axis such that $(Ax)_1 = (Ax)_2 = 0$ and $f^{\omega}(Ax,Av) = f^{\omega}(x,v)$.
But the monotonicity of $\varphi$ implies
\begin{align*}
f^{\omega} (x,v) &=  \varphi (\frac{1}{2} v^2 + U^{\omega} (x) - \frac{1}{2} \omega^2 a^2) = \varphi (E_J(x,v))  \\
&\neq \varphi (E_J(Ax,Av)) = \varphi (\frac{1}{2} v^2 + U^{\omega} (x)) = f^{\omega} (Ax,Av),
\end{align*}
which contradicts our assumption of spherical symmetry. With a similar argument, one can also show that the constructed solutions cannot be axially symmetric with respect to any axis in $\mathbb{R}^3$ except for the $x_3$-axis. Though our deformations only have mirror symmetry with respect to every coordinate plane, which would match a triaxial system, we would still have to prove that the constructed $\zeta ^{\omega}$ are not axially symmetric with repect to the $x_3$-axis to construct triaxial solutions. \\ \\
The asserted continuity properties (iv) can be proved as follows: For $x \in B_3$ we have
$$ |U^{\omega} (x) - U^{\omega '}(x)| \leq \| U_0' \|_{\infty} |g_{\zeta _{\omega}}^{-1} (x) - g_{\zeta _{\omega'}}^{-1} (x)| \leq C \| \zeta _{\omega} - \zeta _{\omega'} \|_X .$$
By the implicit function theorem, $\zeta ^{\omega}$ continuously depends on  $\omega$  with respect to the $\| \cdot \|_X$-norm and we have
$\rho ^{\omega} (x) = \tilde{h} (\omega, r(x) , U^{\omega} (x))$. \\
Lemma \ref{lemma21} implies that $\rho ^{\omega}$ is continuous in $\omega$ with respect to  $\|\cdot \|_{\infty}$ and
$$ U^{\omega} (x) = - \int _{B_3} \frac{\rho ^{\omega} (y)}{|x-y|} dy + U_0 (0) + \int_{B_3} \frac{\rho^{\omega} (y)}{|y|}dy, \quad x \in \mathbb{R}^3 $$
implies the continuity of $U^{\omega}$ in $\omega$ with respect to $\| \cdot \| _{1,\infty}$. Differentiating the above expression for $\rho ^{\omega}$ yields the continuity of $\rho ^{\omega}$ with respect to $\| \cdot \| _{1,\infty}$ and therefore also the continuity of $U^{\omega}$ in the norm $\| \cdot \| _{2,\infty}$. 
\end{proof}
\section{$\partial _{\zeta} T(0,0)$ is an isomorphism} \label{dT2}
In this section, we want to establish some of the assumptions needed for the implicit function theorem.
We will prove the following result:
\begin{proposition} \label{prop41}
The mapping $\partial _{\zeta} T(0,0):X\rightarrow Y$ is a linear isomorphism.
\end{proposition}
Let $\omega_2:= \sqrt{|E_1|}/4$, where $E_1$ is defined in Lemma \ref{lemma21} and let us recall from \cite{REIN}, Proposition 3.1 that the Fr\'echet-derivative of $T:]-\omega_2,\omega _2[ \times \Omega \rightarrow Y$ 
is given by \vspace{5mm}
$$\hspace{-90mm}  [\partial_{\zeta} T(\omega, \zeta) \Lambda] (x) =  $$
\begin{align}
= &- \int_{B_3} \left( \frac{1}{|g_{\zeta} (x)-y|} - \frac{1}{|y|} \right) \partial _u \tilde{h} (\omega, r(y), U_{\zeta} (y)) \nabla U_{\zeta} (y) \cdot \frac{g_{\zeta}^{-1} (y)}{|g_{\zeta}^{-1} (y)|} \Lambda (g_{\zeta}^{-1} (y)) dy \notag \\
&- \int_{B_3} \frac{g_{\zeta} (x) -y}{|g_{\zeta} (x) -y|^3} \tilde{h} (\omega, r(y), U_{\zeta} (y)) dy \cdot \frac{x}{|x|} \Lambda (x), \quad x \in B_4, \label{dert}
\end{align}
\vspace{5mm} where $\omega \in ]-\omega_2,\omega_2[, \, \zeta \in \Omega, \, \Lambda \in X, $ and $U_{\zeta} (y) := U_0 (g_{\zeta}^{-1} (y)), \, \, y \in B_3$ \\
We abbreviate $L_0 \Lambda := \partial _{\zeta} T(0,0)\Lambda$ for $\Lambda \in X$. We observe that $g_0 = id$ and therefore the function $U_{\zeta}$ in (\ref{dert}) coincides with the potential $U_0$ of the spherically symmetric steady state we started with, if $\zeta =0$. We have
\begin{align*}
\rho '_0 (|x|) &= \partial _u\tilde{h} (0,r(x),U_0(|x|))U_0'(|x|) \\
&= \partial _u\tilde{h} (0,r(x),U_0(|x|))\nabla U_0(x)\cdot \frac{x}{|x|}, \quad x \in \mathbb{R}^3.
\end{align*}
This implies
\begin{align*}
(L_0 \Lambda)(x) &= - \int_{B_3} \left( \frac{1}{|x-y|} - \frac{1}{|y|} \right) \rho_0'(|y|) \Lambda (y) \, dy - \int_{B_3} \frac{x-y}{|x-y|^3}\rho_0(|y|)\, dy \cdot \frac{x}{|x|} \Lambda (y) \\
&= - U_0'(|x|)\Lambda (x) - \int_{B_3} \left( \frac{1}{|x-y|} - \frac{1}{|y|} \right) \rho_0'(|y|) \Lambda (y) \, dy, \quad x \in B_4, \, \Lambda \in X.
\end{align*}
Now let
$$ (K\Lambda)(x):= -\frac{1}{U_0'(|x|)} \int_{B_3} \left( \frac{1}{|x-y|} - \frac{1}{|y|} \right) \rho_0'(|y|) \Lambda (y) \, dy, \quad x \in \dot{B}_4, \, \Lambda \in C_S(B_4).$$
Then we can write 
\begin{equation} \label{l0def}
(L_0\Lambda)(x) = -U_0'(|x|)[(id -K)\Lambda](x), \quad x \in B_4,\, \Lambda \in X. 
\end{equation}
In order to prove Proposition \ref{prop41}, we need
\begin{lemma} \label{kompakt}
The linear operator $K:C_S(B_4) \rightarrow C_S(B_4)$ is compact, where $C_S(B_4)$ is equipped with the supremum norm $\|\cdot \|_{\infty}$.
\end{lemma}
\begin{proof}
For $\Lambda \in C_S(B_4)$ let
$$ V_{\Lambda} (x) := - \int_{B_3} \frac{1}{|x-y|} \rho_0'(|y|)\Lambda(y) \, dy, \quad x\in \mathbb{R}^3. $$
Then $V_{\Lambda} \in C^1(\mathbb{R}^3), \enspace \nabla V_{\Lambda} (0)=0$, and
$$ (K\Lambda)(x) = \frac{1}{U_0'(|x|)}(V_{\Lambda} (x) - V_{\Lambda} (0)), \quad x \in \dot{B}_4 .$$
Using Lemma \ref{lemmaueig}(c), we obtain the estimate
$$ |(K\Lambda)(x)| \leq \frac{1}{C|x|} \| \nabla V_{\Lambda} \|_{\infty} |x| \leq C \|\Lambda \|_{\infty}, \quad x \in \dot{B}_4 ,$$
where the constant $C$ depends on $\rho _0$ and $U_0$, but not on $\Lambda$ or $x$. Thus $K$ maps bounded sets into bounded sets. We
next show that $K\Lambda$ is H\"older continuous with exponent 1/2, uniformly on bounded sets in $C_S(B_4)$. Let $M>0$ and assume
$\|\Lambda \|_{\infty} \leq M$. In the following, constants denoted by $C$ depend on $\rho _0, U_0$ and $M$, but not on $\Lambda$. Obviously, $\rho_0' \Lambda \in L^{\infty}(\mathbb{R}^3)$ and we deduce from Lemma \ref{diffu} the existence of $C>0$ with
$$ |\nabla V_{\Lambda} (x) - \nabla V_{\Lambda}(x')| \leq C\| \rho _0'\Lambda\|_{\infty} |x-x'|^{1/2},\quad x,x' \in B_4 $$
Since $\nabla V_{\Lambda} (0) = 0$, the latter implies
$$ |\nabla V_{\Lambda} (x)| \leq C|x|^{1/2}, \quad x\in B_4. $$
Now let $x,x' \in \dot{B}_4$ and $|x|\leq |x'|$. Then
\begin{align*}
|(K\Lambda)(x) - (K\Lambda)(x')| &\leq \bigg| \frac{1}{U_0'(|x|)} - \frac{1}{U_0'(|x'|)} \bigg| \, |V_{\Lambda}(x) - V_{\Lambda} (0)| \\
& \hspace{4mm} +\frac{1}{U_0'(|x'|)}|V_{\Lambda}(x) - V_{\Lambda} (x')| =: I_1 + I_2 
\end{align*}
and we obtain for some $z \in B_4$ with $|z| \leq |x'|$ the estimates
\begin{align*}
I_1 &\leq \frac{|U_0'(|x|) - U_0'(|x'|)|}{|x| \, |x'|}|\nabla V_{\Lambda} (z)|\, |x| \leq C |x-x'|^{1/2} \frac{(|x| + |x'|)^{1/2}}{|x'|} |z|^{1/2} \\
&\leq C|x-x'|^{1/2}, \\
\intertext{and}
I_2 &\leq \frac{C}{|x'|}|\nabla V_{\Lambda} (z)| \, |x-x'| \leq \frac{C}{|x'|} |z|^{1/2} |x-x'| \leq C|x-x'|^{1/2},
\end{align*}
so that
$$ |(K_{\Lambda}) (x) K_{\Lambda}) (x')|\leq C|x-x'|^{1/2}, \quad x,x'\in \dot{B}_4$$
and
$$ |(K_{\Lambda})(x)| \leq  C|\nabla V_{\Lambda} (z)| \leq C|x|^{1/2}, \quad x \in \dot{B}_4.$$
We have shown that $K$ maps bounded sets of $C_S(B_4)$ into bounded and equicontinuous subsets of $C_S(B_4)$. Thus $K$ is compact by the Arzela-Ascoli theorem and the proof is complete. 
\end{proof}
\begin{lemma} \label{Fredholm}
$id -K: C_S(B_4) \rightarrow C_S(B_4)$ is one-to-one and onto.
\end{lemma}
\begin{proof}
Since $K$ is compact, it suffices to show that $id -K$ is one-to-one. Let $\Lambda \in C_S(B_4)$ with $\Lambda - K\Lambda = 0$. Now $\Lambda = 0$ can be shown by expanding $\Lambda$ into spherical harmonics. For that purpose, let
$$\{ \mathcal{S}_{n,j}, \enspace n \in \mathbb{N}, \enspace j=1,\ldots, 2n+1\}$$
be the orthonormal set of spherical harmonics introduced in the Appendix, cf. Section \ref{appendix}, where for $n \in \mathbb{N}$, the functions $\mathcal{S}_{n,j} : \partial B_1 \rightarrow \mathbb{R}, \enspace j=1,\ldots , 2n+1$ are homogeneous polynomials of degree $n$. We define
\begin{equation}
\Lambda _{nj} (r) := \int_{\partial B_1} \mathcal{S}_{n,j} (\xi) \Lambda (r\xi) \, d\omega _{\xi} = \frac{1}{r^2} \int_{\partial B_r} \mathcal{S}_{n,j} (x/r) \Lambda (x) \, d\omega _{x}
\end{equation}
and we use the expansion of the integral kernel $1/|x-y|$ into spherical harmonics, cf. Lemma \ref{newsp1} and Lemma \ref{newsp2}: For $x,y \in \mathbb{R}^3$, $x=r\xi$ and $y= s\eta$ with $\xi, \eta \in \partial B_1$, $r,s\in \mathbb{R}^+,$ $r\neq s$, we have
$$ \frac{1}{|x-y|} = \max(r,s)^{-1}  \sum_{n=0}^{\infty} \sum_{j=1}^{2n+1} \frac{4\pi}{2n+1}\left( \frac{\min(r,s)}{\max(r,s)} \right)^n \mathcal{S}_{n,j}(\xi) \mathcal{S}_{n,j} (\eta).$$
$K\Lambda - \Lambda = 0$ then implies
\begin{align*}
\Lambda _{nj} (r) &= -\frac{1}{U_0'(r)} \int_{B_3} \int_{\partial B_1} \bigg( \frac{1}{|r\xi - y|} - \frac{1}{|y|} \bigg) \mathcal{S}_{n,j} (\xi) \, d\omega _{\xi} \, \rho _0'(|y|) \Lambda (y) \, dy \\
&= -\frac{4\pi}{2n+1}\frac{1}{U_0'(r)}\int_0^3 s^2 \rho _0'(s) \frac{\min(r,s)^n}{\max(r,s)^{n+1}}  \int_{\partial B_1} \mathcal{S}_{n,j} (\eta) \Lambda(s\eta)\, d\omega_{\eta} \, ds \\
& \hspace{7mm} +\frac{4\pi}{2n+1}\frac{1}{U_0'(r)}\int_0^3 s^2 \rho _0'(s) \frac{0^n}{s^{n+1}}  \int_{\partial B_1} \mathcal{S}_{n,j} (\eta) \Lambda(s\eta)\, d\omega_{\eta} \, ds \\
&= -\frac{4\pi}{2n+1}\frac{1}{U_0'(r)}\int_0^3 s^2 \rho _0'(s) \bigg( \frac{\min(r,s)^n}{\max(r,s)^{n+1}} - \frac{0^n}{s^{n+1}} \bigg) \Lambda_{nj} (s) \, ds,
\end{align*}
where we used that the functions $\mathcal{S}_{n,j}$ are orthonormal with repsect to $\langle .,. \rangle _{L^2(\partial B_1)}$. We find that
$$ \Lambda _{01} (r) = - \frac{4\pi}{r U_0'(r)} \int_0^r \rho_0'(s) \, s (s-r) \Lambda _{01}(s) \, ds $$
and we obviously have $\lim_{r \rightarrow 0} \Lambda _{01} (r) = 0$. Let $R \geq 0$ be maximal such that $\Lambda _{01} (r)$ vanishes on $[0,R]$. Then for $r \in [R,3]$,
$$ |\Lambda _{01} (r) | \leq  \frac{4\pi}{r U_0'(r)} \|\rho_0'\|_{\infty} \sup_{0\leq s \leq r}|\Lambda _{01} (s)|\int _R^r s(r-s) \, ds \leq C(r-R)\sup_{0\leq s \leq r}|\Lambda _{01} (s)|. $$
Thus for small $\epsilon >0$, we have $\Lambda _{01}(r) = 0$ on the interval $[R,R+\epsilon]$ and we conclude that $\Lambda _{01}$ vanishes on the whole interval $[0,3]$. Now up to linear combinations, the spherical harmonics for $n=1$ are given by $x_1, x_2, x_3$, and $\Lambda \in C_S$ implies
$$ \int_{\partial B_1} \xi_1 \Lambda (r\xi) \, d\omega _{\xi} = - \int_{\partial B_1} \xi_1 \Lambda (r\xi) \, d\omega _{\xi} = 0,$$
where we made the transformation $\xi \mapsto (-\xi _1, \xi _2, \xi_3)$. Analoguously,
$$ \int_{\partial B_1} \xi_2 \Lambda (r\xi) \, d\omega _{\xi} = \int_{\partial B_1} \xi_3 \Lambda (r\xi) \, d\omega _{\xi} =0, $$
and we have $\Lambda _{11} = \Lambda _{12} = \Lambda _{13} \equiv 0$.
Let $n \geq 2$. Then 
$$ \Lambda _{nj} (r) = -\frac{4\pi}{2n+1}\frac{1}{U_0'(r)} \left( \int_0^r s^2 \rho _0 '(s) \frac{s^n}{r^{n+1}}\Lambda _{nj} (s) \, ds + 
\int_r^3 s^2 \rho _0 '(s) \frac{r^n}{s^{n+1}}\Lambda _{nj} (s) \, ds \right), $$
and 
\begin{align*}
|\Lambda _{nj} (r)| &\leq \frac{4\pi}{2n+1} \frac{1}{U_0'(r)}\|\Lambda _{nj}\|_{\infty} \left(\frac{1}{r^2}\int_0^r (-\rho_0')(s) \frac{s^{n-1}}{r^{n-1}} s^3 \, ds +  r\int_r^3 (-\rho_0')(s) \frac{r^{n-1}}{s^{n-1}} \, ds\right) \\
&\leq \frac{4\pi}{2n+1} \frac{1}{U_0'(r)}\|\Lambda _{nj}\|_{\infty} \left(\frac{1}{r^2}\int_0^r (-\rho_0')(s) s^3 \, ds +  r\int_r^3 (-\rho_0')(s) \, ds\right) \\
&= \frac{4\pi}{2n+1} \frac{1}{U_0'(r)}\|\Lambda _{nj}\|_{\infty} \left(\frac{1}{r^2} r^3 (-\rho_0)(r) + \frac{3}{r^2} \int _0^r s^2 \rho_0 (s) \, ds + r \rho_0(r) \right) \\
&=  \frac{3}{2n+1}\| \Lambda _{nj}\|_{\infty},
\end{align*}
where we integrated by parts in the third line and used the fact that $U_0'(r) = \frac{4\pi}{r^2} \int_0^r s^2 \rho _0(s) \, ds$ in the last line, also 
recall from (\ref{rhoformel}) that $-\rho _0'(r) \geq 0$. \\
Now $2n+1 >3$ for $n\geq 2$ implies that $\Lambda _{nj} \equiv 0$ for $n \geq 2$ as well and the completeness of $\{\mathcal{S}_{n,j}\}$ induces $\Lambda \equiv 0$. We conclude that $id -K$ is one-to-one as claimed.
\end{proof}
It is now clear that $L_0 : X \rightarrow Y$ is one-to-one as well -- this follows from Eq. (\ref{l0def}) and the fact that $U_0'(r) >0$ for $r>0$. So once we have proved the next lemma, the proof of Proposition \ref{prop41} will be complete.
\begin{lemma}
$L_0:X\rightarrow Y$ is onto.
\end{lemma}
\begin{proof}
Let $g \in Y$ and define $q:= g/U_0'$. We will show $q\in X$. We have $q \in C^1(\dot{B}_4) \cap C_S(B_4)$ and
$$ |\nabla q| \leq \frac{|\nabla g(x)|}{U_0'(|x|)} + |g(x)| \enspace \bigg| \frac{U_0''(|x|)}{U_0'(|x|)^2} \frac{x}{|x|} \bigg| \leq C \bigg( \frac{|\nabla g(x)}{|x|} + \frac{|g(x)|}{|x|^2} \bigg) \leq 2C\| g \|_Y.$$
By definition of $Y$ and since $U_0 \in C^2([0,\infty[)$ with $U_0''(0)>0$ we have that for every $x\in \partial B_1$,
\begin{align*}
\nabla q(tx) &= \frac{\nabla g(tx)}{t} \frac{t}{U_0'(t)} - \frac{g(tx)}{t^2} U_0''(t) \bigg( \frac{t}{U_0'(t)} \bigg)^2 x \\
&\rightarrow  \frac{\nabla g(0x)}{0} \frac{1}{U_0''(0)} - \frac{g(0x)}{0^2} U_0''(0)  \frac{1}{U_0''(0)^2}  x
\end{align*}
as $t\rightarrow 0+$, uniformly in $x \in \partial B_1$.

Since $X \subset C_S(B_4)$, there exists by Lemma \ref{Fredholm} an element $\Lambda \in C_S(B_4)$ such that
$$ \Lambda - K\Lambda = -q = -\frac{g}{U_0'}.$$
This implies that $L_0 \Lambda = g$ and thus that $L_0$ is onto, provided $\Lambda \in X$. To see the latter we observe that $\Lambda = K\Lambda - q$ is H\"older continuous since $K\Lambda$  is H\"older continuous. If we now define $V_{\Lambda}$ as above in the proof of Lemma \ref{kompakt} we also conclude that $V_{\Lambda} \in C^2(\mathbb{R}^3)$ and thus $K\Lambda \in C^1(\dot{B}_4)$. Denoting by $H_{V_{\Lambda}}$ the Hessian of $V_{\Lambda}$ we obtain for each $x \in \dot{B}_4$ a point $z \in \overline{0,x}$ such that
\begin{align*}
|\nabla (K\Lambda)(x)| &\leq \bigg| \frac{U_0''(|x|)}{U_0'(|x|)^2} \bigg| \enspace |V_{\Lambda}(x) - V_{\Lambda}(0)| + \frac{1}{|U_0'(|x|)|} |\nabla V_{\Lambda}(x)| \\
&\leq \frac{C}{|x|^2} |\langle H_{V_{\Lambda}}(z)x,x\rangle | + \frac{C}{|x|} |\nabla V_{\Lambda} (x)| \leq C\| D^2 V_{\Lambda} \|_{\infty}
\end{align*}
Finally, for $x\in \partial B_1$, we have
\begin{align*}
\nabla (K\Lambda)(tx) &=  - \frac{U_0''(t)}{U_0'(t)^2}x(V_{\Lambda} (tx) - V_{\Lambda} (0)) + \frac{1}{U_0'(t)}\nabla V_{\Lambda}(tx) \\
&= -U_0''(t) \bigg( \frac{t}{U_0'(t)} \bigg) ^2 x \frac{1}{t^2} \frac{1}{2} \langle H_{V_{\Lambda}} (\tau x) tx,tx \rangle + \frac{t}{U_0'(t)} \frac{\nabla V_{\Lambda} (tx)}{t} \\
&\rightarrow - \frac{1}{2U_0''(0)} \langle H_{V_{\Lambda}} (0)x,x\rangle x + \frac{1}{U_0''(0)}D^2 V_{\Lambda} (0) x,
\end{align*}
as $t\rightarrow 0+$, uniformly in $x \in \partial B_1$. We have shown that $K\Lambda \in X$ and this implies $\Lambda = K\Lambda + q \in X$ and the proof is complete.
\end{proof}
\section{Appendix} \label{appendix}
\setcounter{theorem}{0}
In this section, we firstly state the implicit function theorem which is used for the proof of Theorem \ref{mainth}. Then we give a regularity result for the Poisson equation and finally introduce spherical harmonics and state two important lemmas: an addition theorem and the expansion of the integral kernel $1/|x-y|$ in spherical harmonics.
\begin{theorem} \label{implicit}
Let $X,Y,Z$ be Banach spaces, $U \subset X$ and $V \subset Y$ neighbourhoods of $x_0 \in X$ and $y_0 \in Y$ respectively, $F : U \times V \rightarrow Z$ continuous and continously Fr\'echet-differentiable with respect to the second variable. Suppose also that $F(x_0,y_0) = 0$ and $F_y^{-1} (x_0,y_0) \in \mathfrak{L} (Z,Y)$. \\
Then there exist balls $\overline{B_r}(x_0) \subset U$, $\overline{B_{\delta}} (y_0) \subset V$ and exactly one continuous map $G: B_r(x_0) \rightarrow B_{\delta} (y_0)$ such that $Gx_0 = y_0$ and $F(x,Gx) =  0$ on $B_r (x_0)$.
\end{theorem}
\begin{proof}
\cite{DEIMLING}, Theorem 15.1.
\end{proof}
\begin{lemma} \label{diffu}
Let $n<p\leq \infty$ and let $\rho(x) \in L^p(\mathbb{R}^n)$ with compact support. Define
$$ V_{\rho}(x) := - \int_{\mathbb{R}^n} \frac{1}{|x-y|} \, \rho(y) \, dy $$
Then for every $0 < \alpha < 1 - n/p$ we have $V_{\rho} \in C^{1,\alpha}(\mathbb{R}^n)$ and
$$ |\partial _i V_{\rho}(x) - \partial _i V_{\rho}(x')| \leq C(n,\alpha,p) |x'-x|^{\alpha} \|f\|_p \mathcal{L}^n(\supp \{\rho\} )^{\frac{1-\alpha}{n} - \frac{1}{p}}$$
\end{lemma}
\begin{proof}
\cite{LL}, Theorem 10.2. 
\end{proof}

\noindent
\textbf{Some facts about spherical harmonics} \\ \\
In the following, we use the notation of \cite{SH} and we will always consider the case, where the space dimension $q$ is equal to 3. 
\noindent
For $n \in \mathbb{N}$, consider a homogeneous polynomial $H_n$ of degree $n$, which satisfies
$$ \Delta H_n (x) = 0.$$
Then for $\xi \in \partial B_1:= \{ x \in \mathbb{R}^3 \, | \, |x| =1 \}$,
$$ S_n (\xi ) := H_n (\xi) $$
is called a spherical harmonic of order $n$. For each $n$, there exist $2n+1$ linearly independent spherical harmonics, which we call $S_{n,j}, \enspace j=1,\ldots 2n+1$, cf. \cite{SH}, Lemma 4. 
\noindent
We denote by $\{ \mathcal{S}_{n,j}, \enspace n=0,\ldots ,\infty, \enspace j= 1, \ldots, 2n+1\} $ the orthonormal set of all spherical harmonics, where we orthonormalize with respect to $\langle .,. \rangle _{L^2(\partial B_1)}$. Then we have the following
\begin{lemma} \label{newsp1}
For a fixed $n \in \mathbb{N}$ and $\xi, \, \eta \in \partial B_1$, we have
$$ \sum_{j=1}^{2n+1} \mathcal{S}_{n,j}(\xi) \mathcal{S}_{n,j} (\eta) = \frac{2n+1}{4\pi} P_n(\xi \cdot \eta), $$
where $P_n(t)$ is the Legendre Polynomial of degree $n$.
\end{lemma}
\begin{lemma} \label{newsp2}
Let $x,y \in \mathbb{R}^3$ with $x= R\xi, \enspace y= r\eta$, for suitable $\xi, \eta \in \partial B_1$ and $r,R \in \mathbb{R}$. Then we have for $R>r$
$$ \frac{1}{|x-y|} = R^{-1} \sum_{n=0}^{\infty} \left( \frac{r}{R} \right)^n P_n(\xi \cdot \eta),$$
and for $R<r$
$$ \frac{1}{|x-y|} = r^{-1} \sum_{n=0}^{\infty} \left( \frac{R}{r} \right)^n P_n(\xi \cdot \eta),$$
where $P_n(t)$ is the Legendre Polynomial of degree $n$.
\end{lemma}
\noindent
Proofs can be found in \cite{SH}, Theorem 2 and Lemma 19. \\ \\
\textbf{Acknowledgements.}
The author wishes to thank Gerhard Rein for the critical review of the manuscript. This research was supported by the Deutsche Forschungsgemeinschaft under the project ``Nichtlineare Stabilit\"at bei ki\-ne\-ti\-schen Modellen aus der Astrophysik und Plasmaphysik''.
\bibliography{galaxy_jac}{}
\bibliographystyle{plain}
\end{document}